\documentclass[a4paper,11pt]{article}
\usepackage{aaskaiid}
\usepackage{orcidlink}
\usepackage{upgreek}

\def \deg         {\text{$^{\circ}$}}

\def \arcsec      {\text{$^{\prime\prime}$}}

\title{Tracing the Cosmic web across Cosmic time through SKA observations of radio galaxies}
\ShortTitle{SKA radio galaxies as tracers of the Cosmic web}

\author[1]{Pratik Dabhade\orcidlink{0000-0001-9212-3574}}
\ShortName{Dabhade et al.} % shortened name list for header 
\author[2]{Gianluca Castignani\orcidlink{0000-0001-6831-0687}}
\author[3]{Mousumi Mahato\orcidlink{0000-0002-6658-7559}}
\author[3]{Shishir Sankhyayan\orcidlink{0000-0003-2601-2707}}
\author[4,5]{D.J. Saikia\orcidlink{0000-0002-4464-8023}}
\author[6]{Viviana Casasola\orcidlink{0000-0002-3879-6038}}
\author[7]{Francoise Combes\orcidlink{0000-0003-2658-7893}}
\affiliation[1]{Astrophysics Division, National Centre for Nuclear Research, Pasteura 7, 02-093 Warsaw, Poland}
\emailAdd{pratik.dabhade@ncbj.gov.pl}
\affiliation[2]{INAF-Osservatorio di Astrofisica e Scienza dello Spazio di
Bologna, Via Piero Gobetti 93/3, 40129 Bologna, Italy}
\emailAdd{gianluca.castignani@inaf.it}
\affiliation[3]{Tartu Observatory, University of Tartu, Observatooriumi~1, 61602 T\~oravere, Estonia}
\affiliation[4]{Fakult\"at f\"ur Physik, Universit\"at Bielefeld, Postfach 100131, D-33501 Bielefeld, Germany}
\affiliation[5]{Assam Don Bosco University, Tapesia, Guwahati 781017, India  }
\affiliation[6]{INAF - Istituto di Radioastronomia, Via Gobetti 101, I-40129 Bologna, Italy}
\affiliation[7]{Observatoire de Paris, LUX, Collège de France, CNRS, PSL University, Sorbonne University, 75014, Paris, France}

\abstract{The Square Kilometre Array will transform studies of the cosmic web by tracing radio galaxies (RGs) and star-forming systems across cosmic time with unprecedented sensitivity, angular resolution, frequency coverage, and survey speed. Powered by accreting supermassive black holes, RGs are not only signposts of AGN feedback but also incisive probes of their environments, from dense clusters to the low-density intergalactic medium. Their lobes, magnetic fields, and energy outflows encode the thermal and non-thermal histories of the surrounding gas, offering diagnostics of IGM pressure, particle ageing, and magnetisation over megaparsec scales.
With its broad frequency coverage (50 MHz–15 GHz), \(\upmu\mathrm{Jy}\) to sub-\(\upmu\mathrm{Jy}\) continuum sensitivity, and wide field of view, the SKA will detect vast radio-source populations across broad ranges of redshift and environment. Measurements of source size, morphology, spectral ageing, radio power, polarisation, and Faraday rotation will reveal how the environment regulates jet propagation and lobe evolution, how radio plasma heats and magnetises the intracluster and intergalactic media, and how early AGN activity influences galaxy growth and star formation in protoclusters. Combined with host identifications, spectroscopic redshifts, and optical, infrared, X-ray, Sunyaev--Zel'dovich, and cosmic-web catalogues, SKA observations will place RGs within their three-dimensional large-scale environments. This chapter presents a framework for using RGs to trace and probe the cosmic web, from nearby filaments and clusters to high-redshift protoclusters, and to test how environment, magnetic fields, feedback, and gas dynamics shape radio-galaxy evolution, protocluster assembly, and star formation across cosmic time.}

\begin{document}
\maketitle

\newpage  % Start the main text on a new page

\section{Introduction: Radio Galaxies as probes of the Cosmic Web in the SKA era}
\vspace{-0.35cm}
Radio galaxies (RGs) are among the most powerful manifestations of black-hole activity in the Universe. Their jets and lobes transport energy, particles, and magnetic fields from active galactic nuclei (AGN) into the surrounding interstellar, circumgalactic, and intergalactic media. Because these structures can extend from kiloparsec to megaparsec scales, they are sensitive to the density, pressure, dynamics, and magnetisation of the environments through which they propagate. RGs, therefore, provide a natural link between the physics of active galactic nuclei and the large-scale cosmic web.

The Square Kilometre Array will transform this field by enabling radio galaxies to be studied not only as individual systems, but as statistical probes of cosmic-web environments across cosmic time. Its combination of low-frequency sensitivity, angular resolution, wide-area survey capability, and polarimetric performance will allow radio morphology, spectral ageing, diffuse emission, and Faraday rotation to be connected with the underlying distribution of galaxies, groups, clusters, filaments, and protoclusters. This chapter outlines how SKA observations of radio galaxies can be used to trace the structure, evolution, and magnetisation of the cosmic web.

\subsection{Radio Galaxies}
\vspace{-0.2cm}
RGs represent a major class of AGN in which accretion onto a supermassive black hole (SMBH) drives the formation of powerful, collimated jets that emit synchrotron radiation across the electromagnetic spectrum. The basic structure of an RG encompasses a compact radio core, often associated with the SMBH and accretion disc, bipolar relativistic jets, and extended radio lobes formed where the jets interact with the surrounding interstellar medium (ISM) and intergalactic medium (IGM). These systems span physical scales from parsecs to several megaparsecs and radiative powers in the range \(L_{\mathrm{rad}} \sim 10^{38}\)--\(10^{46}\) erg s\(^{-1}\) \citep{Blandford2019,Hardcastle2020,Saikia2022,Dabhade2023}.

The identification of double-lobed radio sources such as Cygnus~A and the discovery of quasars established that powerful radio emission is produced by synchrotron-emitting plasma associated with accreting supermassive black holes \citep{Jennison1953,Baade1954,Burbidge1956,Schmidt1963}. The morphological dichotomy introduced by \citet{Fanaroff1974} distinguishes two principal types: FR\,I sources with edge-darkened, decelerating jets, and FR\,II sources with edge-brightened lobes and compact hotspots. This division reflects differences in jet kinetic power and environmental interaction \citep{Begelman1984, Laing2014}.  

Orientation-based unification explains much of the observed diversity among intrinsically related radio-loud AGN \citep{Urry1995}. Jets viewed close to the line of sight appear as Doppler-boosted blazars, whereas larger viewing angles reveal extended jets and lobes. However, excitation state, radio morphology, jet power, and evolution also depend on accretion mode, host properties, and environment.

The launching and collimation of these jets are understood as magnetohydrodynamic processes that extract rotational energy from either the accretion disc or the spinning black hole itself \citep{Blandford1977, Blandford1982}. The jets transport energy, momentum, and magnetic flux from the black-hole environment to extragalactic scales, inflating radio lobes and heating the surrounding medium. In this way, RGs act as a major channel of AGN feedback, regulating gas cooling, star formation, and the thermodynamic state of their host environments \citep{McNamara2007}. Their massive elliptical hosts and extended hot haloes favour recurrent accretion and sustained jet activity, processes that were more prevalent at earlier cosmic epochs when massive galaxies and black holes were rapidly growing. The relative abundance of powerful RGs at high redshift (high-\(z\)) implies that their jets contributed significantly to the global energy budget of the Universe and played a vital role in shaping the thermodynamic and magnetic properties of the cosmic environment \citep{Blandford2019}.

Because their extended radio structures preserve signatures of jet propagation, particle ageing, recurrent activity, and interaction with the ambient medium, RGs provide valuable diagnostics of AGN duty cycle and jet--environment coupling \citep[see also ][]{Hardcastle01.2026.SKA}.

\subsection{Rotation Measure Grids}\label{sec:rmgrid}
\vspace{-0.2cm}
A \textit{Rotation Measure (RM) grid} provides a quantitative framework to investigate magnetic fields in the extragalactic Universe through Faraday rotation of linearly polarised radio emission from background and embedded sources such as RGs and quasars. Each RM value encapsulates the integrated line-of-sight effect of magnetised plasma, expressed as  
\begin{equation}
\mathrm{RM} = 0.812 \int_0^L n_e(l)\, B_\parallel(l)\, \mathrm{d}l \quad [\mathrm{rad\, m^{-2}}],
\end{equation}
where $n_e$ is the thermal electron density in cm$^{-3}$, $B_\parallel$ is the magnetic field component parallel to the line of sight in $\upmu$G, $l$ is the path length in parsecs, and \(L\) is the path length through the magneto-ionic medium. The observed polarisation angle $\chi$ of synchrotron emission varies with the square of the wavelength as  
% \begin{equation}
$\chi(\lambda) = \chi_0 + \mathrm{RM}\,\lambda^2$,
% \end{equation}
where $\chi_0$ represents the intrinsic polarisation angle of the source \citep[see][~for a review]{Saikia1988}. Multi-frequency polarimetric observations allow this dependence to be measured and fitted, yielding an RM for each polarised source. When these RMs are sampled over a large number of extragalactic sources, they form a grid that traces the magneto-ionic medium across cosmic structures.

In the extragalactic context, RM grids are a powerful diagnostic of magnetism in a variety of environments: from RGs and clusters to filaments of the cosmic web and the IGM. The first large-scale RM compilations (e.g., \citealt{Simard-Normandin1981, Kronberg1982, Taylor2009}) established the feasibility of using background sources to map foreground magnetic structures. Subsequent studies have demonstrated their application in probing cluster magnetic fields \citep[e.g.][]{Govoni2004, Bonafede2010}, radio galaxy environments \citep{Laing1988, Guidetti2011, Sullivan2018}, and the magnetised cosmic web \citep{Akahori2011, Vacca2018, Vernstrom2021}. These developments have opened the way to using RM grids not only for individual objects, but also for statistical studies of cosmic magnetism through cross-correlations with galaxy density fields, clusters, filaments, and large-scale structure catalogues.

Broadband polarisation data also allow \textit{RM synthesis} \citep{Brentjens2005} and \textit{Faraday tomography} to recover the Faraday dispersion function $F(\phi)$, providing the full distribution of polarised emission as a function of Faraday depth $\phi$. This approach enables the identification of multiple magneto-ionic components along the line of sight and helps distinguish between internal and external Faraday rotation. Depolarisation effects and Faraday complexity, which vary with wavelength and environment, thus become key diagnostics of magnetic field geometry, turbulence, and thermal plasma content in and around RGs \citep[e.g.][]{Anderson2018, Sullivan2018}.

Recent wide-band polarisation surveys such as LOFAR, ASKAP-POSSUM, and MeerKAT \citep[e.g.][]{Anderson2021, Hutschenreuter2022,Sullivan2023} have dramatically increased both the precision and surface density of RM measurements, enabling detailed magneto-ionic mapping across large extragalactic volumes \citep[see also ][]{OSullivan01.2026.SKA}. RM grids derived from these surveys reveal magnetic field amplification and structure in cluster outskirts, filamentary bridges, and the lobes of RGs, providing new insights into environmental asymmetries, entrainment, and feedback processes. 
A major recent advance is the second SPICE-RACS data release, which provides the largest single RM catalogue produced to date. Covering \(\sim88\%\) of the sky, it contains \(\sim2.5\times10^{5}\) reliable RMs at \(>8\sigma\), with a median uncertainty of \(\sim2\,\mathrm{rad\,m^{-2}}\), an areal density of \(\sim6.7\,\mathrm{deg^{-2}}\), and an effective RM-grid spacing of \(\sim23^{\prime}\) \citep{Thomson2026SPICE-RACS}. This result demonstrates that wide-area RM-grid science has already entered the statistical regime and provides an important observational benchmark for forthcoming POSSUM and SKA polarisation surveys.

In the SKA era, RM grids will undergo a transformative increase in sensitivity and source density. Deep polarisation surveys, particularly with SKA1-Mid, are expected to reach RM densities of up to several thousand sources per square degree in the deepest fields, although the achievable density will depend on observing frequency, angular resolution, polarised-source counts, Faraday complexity, and depolarisation \citep{Beck2015,Johnston-Hollitt2015}. These grids will enable statistical measurements of magnetic-field amplification and ordering in RGs, clusters, and the cosmic web, including RM-excess measurements and cross-correlations with foreground filaments \citep{Vazza2021,Sullivan2023}. Combined with optical, X-ray, and Sunyaev--Zel'dovich observations and independent constraints on electron density, they will support the tomographic and statistical characterisation of magnetised plasma and provide stronger constraints on primordial and intergalactic magnetic-field models \citep{Durrer2013,Carretti2022}.
Together, these data will enable population-level studies of magnetised feedback in AGN jets and lobes, quantify how large-scale environments influence radio-galaxy morphology, and constrain the redshift evolution of cosmic magnetism.

\subsection{Radio Galaxies and RM Grids}\label{sec:rgrm}
\vspace{-0.2cm}
The propagation of jets into the surrounding IGM naturally connects the physics of RGs to the structure and magnetisation of the cosmic web.
RGs play a dual role in the study of cosmic magnetism, both benefiting from and contributing to RM grids. 

\textit{(i) Benefiting from RM grids.}  
RM grids provide essential foreground and environmental context for interpreting Faraday rotation in RGs. The RM of an RG includes contributions from the Milky Way, the IGM, its surrounding environment, and the galaxy itself. Dense RM grids allow the Galactic and large-scale foreground components to be modelled and subtracted more accurately, helping to isolate the intrinsic and environmental Faraday rotation associated with the source. This facilitates precise measurements of magnetic field strength, ordering, and thermal plasma density in radio lobes, jets, and halos, and helps identify asymmetries arising from jet–medium interactions or environmental confinement \citep[e.g.][]{Laing1988, Guidetti2011,Sullivan2018}. In the \textit{SKA era}, with RM grid densities exceeding $10^3$~deg$^{-2}$, it will be possible to map foreground magneto-ionic structures at arcminute resolution, refining 3D models of the local Faraday environment around individual sources.

\textit{(ii) Contributing to RM grids.}  
RGs themselves constitute a major component of RM grids. Their bright, polarised synchrotron emission provides numerous, reliable sightlines that sample the magneto-ionic medium across diverse extragalactic environments. Each sufficiently resolved extended source can contribute multiple RM measurements across its lobes and hotspots, improving the statistical sampling of both Galactic and intergalactic magnetic fields. At low frequencies, faint polarised RGs significantly enhance RM grid density, supporting studies of the cosmic web and large-scale magnetogenesis \citep[e.g.][]{Akahori2011, Vazza2021, Vernstrom2021}.

In the SKA era, RM-grid densities approaching or exceeding \(10^{3}\,\mathrm{deg^{-2}}\) in deep fields will substantially improve foreground modelling and the statistical characterisation of local Faraday structure around individual RGs, although the effective angular resolution will depend on source density, interpolation method, and foreground complexity.
Extended RGs can contribute multiple RM measurements across their lobes and hotspots, but the number of independent sightlines and their low-frequency contribution will depend on angular resolution, Faraday complexity, and depolarisation. As contributors, RGs will add numerous well-characterised sightlines to RM grids; as beneficiaries, they will be studied with greatly improved Faraday-depth information.

\begin{figure}
    \centering
	\includegraphics[width=\linewidth]{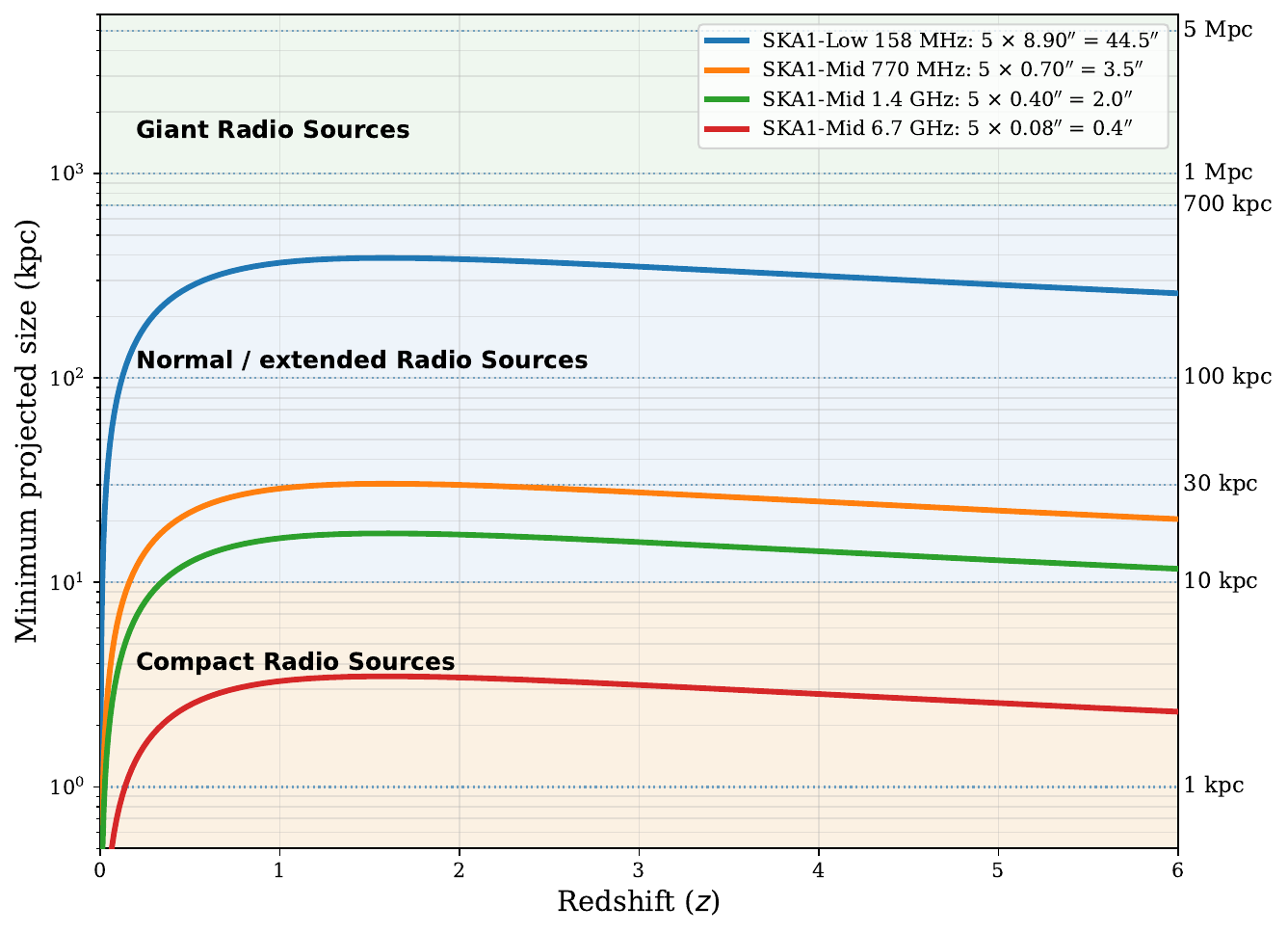}
    \caption{
Angular-resolution limits for morphological identification of radio galaxies/quasars with representative SKA1 observing bands.
The curves show the minimum projected linear size required for a source to span five synthesised beams, adopted here as a
practical criterion for recognising extended or double-lobed radio morphology. The calculation uses the angular-diameter distance 
as a function of redshift and representative SKA1 beam sizes: \(8.9^{\prime\prime}\) for SKA1-Low at 158\,MHz, \(0.7^{\prime\prime}\) for
SKA1-Mid at 770\,MHz, \(0.4^{\prime\prime}\) for SKA1-Mid at 1.4\,GHz, and \(0.08^{\prime\prime}\) for SKA1-Mid at 6.7\,GHz \citep{braun2019anticipatedperformancesquarekilometre}. 
The shaded horizontal regions indicate approximate source-size regimes: compact radio sources below \(\sim10\)~kpc, normal or extended radio sources between \(\sim10\) and \(700\)~kpc,
and giant radio sources above \(\sim700\)~kpc. Sources lying above a given curve are large enough, in angular-resolution terms, to be morphologically resolved in that band,
 whereas sources below the curve would require higher angular resolution for reliable structural classification. This figure illustrates that SKA1-Mid at 1.4\,GHz can resolve tens-of-kpc-scale radio galaxies over a large fraction of cosmic time, while higher-frequency SKA1-Mid observations probe compact radio structures, and SKA1-Low is primarily suited to large, diffuse, steep-spectrum systems. These limits describe angular resolvability only; detectability will also depend on surface brightness, signal-to-noise ratio, spectral shape, and imaging strategy. The angular-to-linear conversion uses the angular-diameter distance derived from the comoving distance at each redshift, assuming a flat \(\Lambda\)CDM cosmology with \(H_{0}=67.4~\mathrm{km~s^{-1}~Mpc^{-1}}\), \(\Omega_{\mathrm{m}}=0.315\), and \(\Omega_{\Lambda}=0.685\) \citep{Planck2020}.}
\label{fig:ska_morph_resolvability}
\end{figure}

\vspace{-0.2cm}
\subsection{SKA Observational Framework and Working Assumptions}
\label{sec:ska_framework}
\vspace{-0.2cm}
The science cases discussed in this chapter are anchored in the complementary capabilities of SKA1-Low and SKA1-Mid. SKA1-Mid, operating over 0.35--15~GHz, will provide sub-arcsecond imaging at GHz frequencies, reaching ($\sim0.4^{\prime\prime}$) resolution at 1.4~GHz and continuum sensitivities of ($\sigma_{\mathrm{cont}}\approx2~\upmu\mathrm{Jy~beam^{-1}}$) in 1~hr \citep{braun2019anticipatedperformancesquarekilometre}. SKA1-Low, operating over 50--350~MHz, will provide the low-frequency sensitivity needed to detect steep-spectrum and low-surface-brightness synchrotron plasma, with representative performance of ($\sim4^{\prime\prime}$) resolution at 300~MHz and ($\sim14~\upmu\mathrm{Jy~beam^{-1}}$) continuum sensitivity in 1~hr \citep{braun2019anticipatedperformancesquarekilometre}. These values are representative and depend on observing strategy, weighting, bandwidth, sky brightness, and final deployment assumptions.

These expectations are supported by SKA-era radio-sky simulations. The SKADS/S3-SEX simulation models \(\sim3.2\times10^{8}\) extragalactic radio sources down to nanojansky flux densities at five frequencies between 151~MHz and 18~GHz, including star-forming galaxies, radio-quiet AGN, and radio-loud AGN populations associated with FR~I/FR~II systems \citep{Wilman2008SKADS}. T-RECS extends this framework over 150~MHz--20~GHz by including AGN/SFG populations, polarisation, and realistic clustering based on dark-matter haloes \citep{Bonaldi2019TRECS}. Together, these simulations provide a quantitative basis for SKA source-yield forecasts, while remaining sensitive to assumptions about luminosity-function evolution and source classification.

The practical consequence of these angular resolutions is shown in Fig.~\ref{fig:ska_morph_resolvability}, which translates representative SKA1 beam sizes into the minimum projected source sizes required for morphological recognition (resolving their cores, jets, lobes, and compact structures). It highlights the complementary roles of SKA1-Mid in resolving compact and normal radio-galaxy structures and SKA1-Low in detecting larger, diffuse, steep-spectrum systems.

For RG studies, the relevant SKA observables include total-intensity morphology, angular and projected linear size, arm-length ratio, lobe axial ratio, core dominance, misalignment angle, integrated radio luminosity, spectral index, spectral curvature, polarised emission, and Faraday rotation measure. SKA will allow RGs to be characterised not only by their luminosity and morphology, but also by their spectral state, polarisation properties, and interaction with the surrounding medium.

The interpretation of these radio observables requires environmental information from optical, infrared, X-ray, and SZ surveys. In particular, spectroscopic and photometric surveys such as DESI, 4MOST, \textit{Euclid}, and LSST will provide the redshift and large-scale-structure context needed to locate radio galaxies within clusters, groups, filaments, sheets, and voids. Throughout this chapter, we therefore treat SKA observations and multi-wavelength environmental mapping as a combined framework for studying how RGs trace, respond to, and influence the cosmic web.

\vspace{-0.3cm}
\section{The Cosmic Web and the Role of Radio Galaxies}
\vspace{-0.35cm}
\subsection{The cosmic web}
\vspace{-0.25cm}
The cosmic web is the fundamental large-scale structure of the Universe, arising from the anisotropic gravitational collapse of matter in the $\Lambda$CDM framework \citep{Zeldovich1970, Bond1996}. It manifests as a vast network of clusters, filaments, sheets, and voids that collectively trace the hierarchical assembly of cosmic matter \citep{vandeWeygaert2008, Cautun2014}. Filaments serve as bridges channelling gas, galaxies, and dark matter between dense nodes, while voids dominate the volume but contain little mass. Cosmological simulations indicate that filaments contain a substantial fraction of the matter in the Universe, often of order 40-50\%, while occupying only $\sim$6\% of the cosmic volume, although the inferred fractions depend on the web-classification method and smoothing scale \citep{Cautun2014,OKane2024}. Their transverse dimensions span a broad range, with prominent low-redshift filaments commonly having widths of order 1–2 Mpc, and becoming progressively thicker and more massive towards the present epoch \citep{Wang2024}.

The morphology and connectivity of the web emerge from the large-scale tidal field, which induces sequential collapse along preferred directions, first forming walls, then filaments, and finally clusters \citep[e.g.,][]{Arnold1982, Shandarin1989}. Quantitative descriptions of this multiscale, topologically complex network rely on advanced structure identification techniques such as {\sc DisPerSE} \citep{Sousbie2011} and {\sc NEXUS} \citep{Cautun2013}, enabling consistent detection of web components in both simulations and galaxy surveys \citep[e.g.][]{Tempel2014,Castignani2022_catalog}. Observationally, the cosmic web has been traced across multiple wavelengths and surveys such as SDSS and LOFAR, revealing that galaxies within filaments differ systematically from those in the field in colour, morphology, stellar mass, and gas content \citep[e.g.,][]{Malavasi2017,Kraljic2018,CroneOdekon2018,Laigle2018, Sarron2019,Castignani2022_VirgoFilGas,Hoosain2024,LylaJung2025}. By studying the large-scale filaments surrounding the Virgo cluster, the benchmark system in the local Universe, \citet{Castignani2022_VirgoFilGas} found clear evidence for a progression from field to filament and cluster galaxies, characterised by decreasing star formation rates, an increasing fraction of galaxies in the quenching phase, a higher proportion of early-type galaxies, and reduced gas content.

\begin{figure}
    \centering
    \includegraphics[width=1\linewidth]{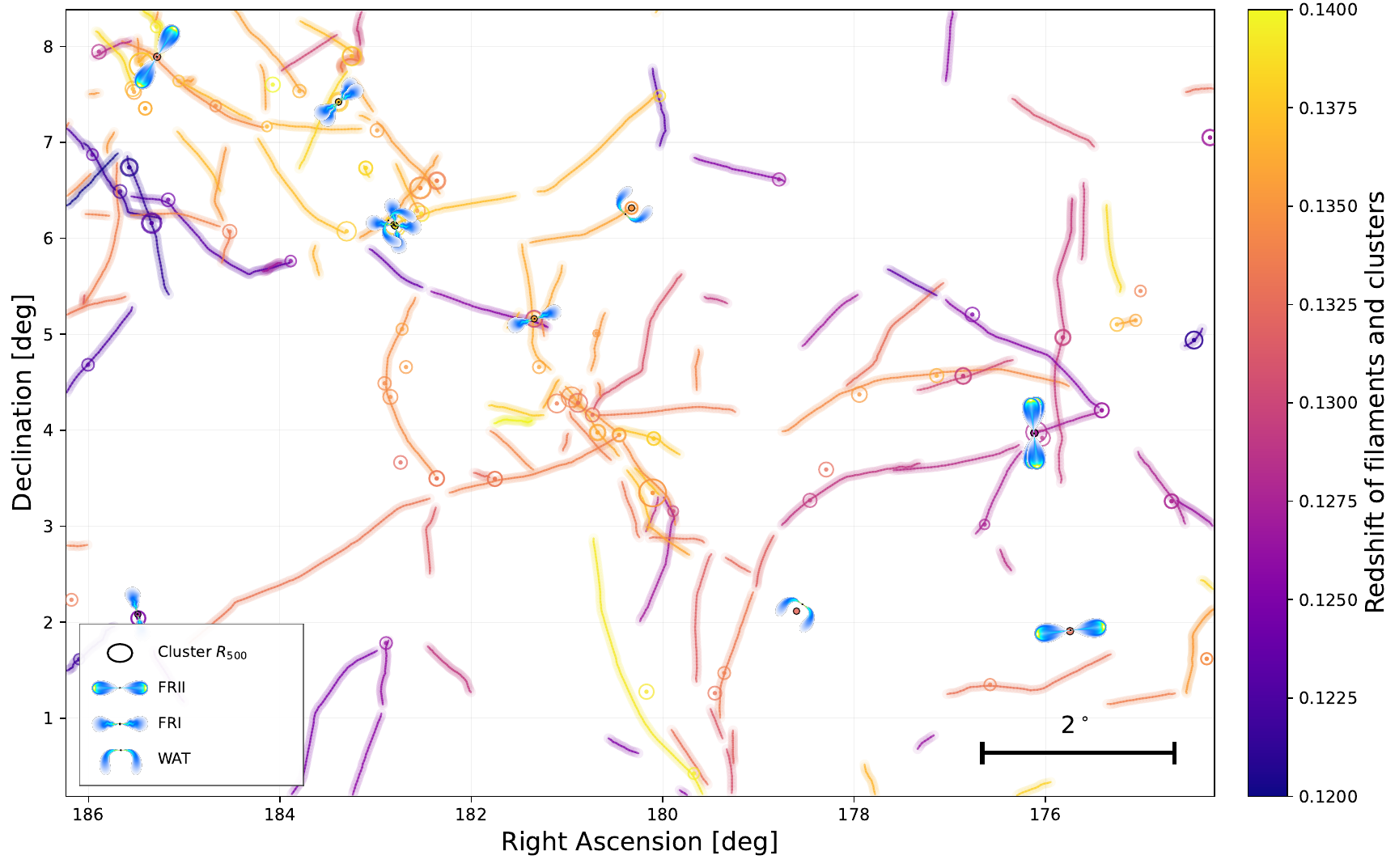}
   \caption{Projected view of the low-redshift cosmic web, covering \(12^\circ\times8^\circ\) within \(0.120\leq z<0.140\), designed to illustrate the locations of radio galaxies with different morphologies relative to the surrounding large-scale structure. Filament spines from \citet{Tempel2014} are plotted at their catalogue sky positions and colour-coded by median filament redshift. Their projected extents follow the catalogue geometry, while the displayed line widths are illustrative and do not represent physical filament widths. Galaxy clusters from \citet{Wen2015,Wen2024} are shown as circles corresponding to \(R_{500}\) and are also colour-coded by redshift. FR\,I, FR\,II, and wide-angle-tail radio galaxies are compiled from the \citet{Capetti2017,Capetti2017FRi}, \citet{Mahato2025_SAGANVI}, \citet{Goyal2026}, and \citet{Missaglia2019} catalogues and represented by schematic morphology icons, with the surrounding haloes indicating source redshift. The icon orientations are randomised for visual clarity and do not represent measured radio position angles, while their plotted sizes are not proportional to the true angular extents of the sources. This two-dimensional representation places different radio-galaxy morphologies within their broader cosmic-web environments. The redshift selection supports genuine three-dimensional proximity, although detailed membership and interaction require source-by-source analysis; the sample is not statistically complete or homogeneous. The cosmological parameters adopted here are the same as those used in Fig.~\ref{fig:ska_morph_resolvability}.} \label{fig:rg_cosmic_web_field}
\end{figure}

These findings show that the cosmic web is not only the geometric framework of large-scale structure, but also the dynamical environment within which galaxies evolve. Environmental processes become progressively stronger from low-density fields to filaments, groups, and clusters, affecting gas content, star formation, morphology, and nuclear activity \citep[e.g.,][]{Kenney2014,Poggianti2017,Castignani2018,Castignani2020_z17,Castignani2020_LIRGs,Castignani2025,Zhang2021,Borrow2023}. For radio galaxies, this environmental dependence is particularly important because jets and lobes propagate through the same gaseous structures that define the cosmic web. Their morphology, size, spectral ageing, polarisation, and Faraday rotation can therefore encode information about the density, pressure, turbulence, and magnetisation of the surrounding medium.

In this context, SKA observations will provide a direct radio view of how AGN activity is connected to the large-scale environment. SKA1-Low will be sensitive to steep-spectrum and low-surface-brightness plasma from extended and remnant radio galaxies, while SKA1-Mid will provide the angular resolution and polarimetric capability needed to resolve jets, lobes, and magneto-ionic structure. Combined with cosmic-web reconstructions from spectroscopic surveys, these observations will allow radio galaxies to be studied as a function of their location within filaments, nodes, groups, clusters, and lower-density regions (see Fig.~\ref{fig:rg_cosmic_web_field}).

% \subsection{Spectroscopic mapping of  the cosmic web}
\subsection{From Redshifts to Cosmic-Web Maps for SKA Science}
\vspace{-0.2cm}
Radio data by themselves cannot establish the full three-dimensional environment of a radio galaxy; without reliable redshift information, even detailed radio morphology remains vulnerable to projection effects and ambiguous environmental associations. Recent SKA pathfinder results quantify this requirement with updated wide-field data. In the EMU pilot RG-CAT catalogue, \(73\%\) of radio sources have CatWISE counterparts, but only \(36\%\) have photometric redshifts, illustrating that host identification, redshift assignment, and environmental classification remain separate steps even in modern multi-wavelength catalogues \citep{Gupta2024RGCAT}. Even in LoTSS DR2, one of the deepest and most sensitive wide-area low-frequency radio surveys to date, $\sim$\,1.8 million sources are detected above $S_{144} \geq 1.1$ mJy across $\sim 5200~{\rm deg}^{2}$. However, only about half of these have an optical identification together with a usable redshift estimate, and spectroscopic information is available for only a small minority \citep{Hardcastle2025}. Spectroscopic and high-quality photometric redshifts are therefore essential for placing SKA-detected radio sources within the cosmic web, distinguishing clusters, groups, filaments, sheets, and voids, and separating physically associated structures from projected alignments. For SKA studies, such redshift information provides the environmental reference frame within which radio morphology, spectral properties, polarisation, and Faraday rotation can be interpreted.

SDSS provided the foundational spectroscopic map of the low-redshift Universe, enabling the first large statistical catalogues of groups, clusters, filaments, and voids. The next generation of surveys will extend this framework by orders of magnitude. DESI has already released spectra for 18.7 million objects in DR1, including more than 13 million galaxies and 1.5 million quasars, and has observed more than 47 million galaxies and quasars as part of its full survey programme \citep{DESI2024}. Its target classes span bright galaxies at low redshift, luminous red galaxies, emission-line galaxies, quasars, and Ly\(\alpha\)-forest tracers, providing a dense map of large-scale structure from the nearby Universe to \(z>2\). The \textit{4-metre Multi-Object Spectroscopic Telescope} \citep[4MOST;][]{de_Jong19} will provide a complementary southern-hemisphere spectroscopic framework, with the Cosmology Redshift Survey designed to obtain millions of redshifts over several thousand square degrees and to trace large-scale structure over \(z\simeq0.15\)--3.5 \citep{Richard20194most}. WEAVE will add highly multiplexed spectroscopy in the northern sky, including galaxy evolution, cluster, and quasar/IGM surveys that are particularly relevant for mapping environments around radio sources. MOONS will extend this spectroscopic framework to higher redshifts. Its near-infrared multi-object capability over a \(\sim500~\mathrm{arcmin^2}\) \citep{Cirasuolo2020} field will detect rest-frame optical lines such as H\(\alpha\), H\(\beta\), [O{\sc ii}], and [O{\sc iii}], enabling spectroscopic mapping of \(z\gtrsim2\) protoclusters and associated filamentary structures through surveys such as MOONRISE \citep{Maiolino2020}.

Space- and ground-based imaging surveys will provide an equally important layer of environmental information. \textit{Euclid} will survey more than \(15,000~\mathrm{deg}^{2}\) of the extragalactic sky with optical imaging, near-infrared photometry, and near-infrared spectroscopy, yielding imaging, photometric, and spectroscopic information for millions of galaxies \citep{Adam2019}. LSST will provide deep multi-epoch optical imaging over a similarly large southern footprint, enabling photometric redshifts, host-galaxy identification, variability information, and weak-lensing measurements for very large galaxy samples. Together, \textit{Euclid}, LSST, DESI, 4MOST, and WEAVE will provide the redshift, stellar-mass, host-galaxy, and density-field context needed to interpret SKA radio sources as tracers of large-scale structure.

These surveys will also support the construction of the cosmic-web catalogues needed for SKA cross-correlation studies. Groups and clusters can be identified through spectroscopic membership, halo-based catalogues, red-sequence methods, X-ray emission, or SZ-selected samples. Beyond collapsed haloes, the filamentary and void network can be characterised using a range of complementary algorithms, including density-ridge and skeleton methods, marked point-process models such as the Bisous formalism, tidal-tensor classifications, graph-based approaches, watershed void finders, and spherical-under-density techniques \citep[e.g.,][]{Tempel2014, Cautun2014, Nadathur2016}. Since each method encodes a different operational definition of the cosmic web, SKA-based environmental studies will need to account for catalogue selection, smoothing scale, redshift completeness, and survey geometry.

By the time SKA Phase~1 reaches maturity, these surveys will provide much of the redshift and environmental scaffolding required for SKA cosmic-web studies. DESI and WEAVE will provide dense northern spectroscopic coverage, while 4MOST will deliver complementary southern mapping, together linking SKA survey fields to large-scale-structure reconstructions over a broad redshift range. \textit{Euclid} and LSST will add wide-area imaging, photometric redshifts, weak-lensing information, host-galaxy properties, and high-redshift environmental constraints, including in regions where spectroscopy remains incomplete. In combination with X-ray and SZ-selected cluster samples, these data sets will allow SKA continuum, spectral-index, polarisation, RM-grid, and {\sc Hi} measurements to be interpreted in terms of physical environment rather than projected sky position alone. This multi-wavelength framework will move radio-galaxy environmental studies from individual systems to statistically controlled samples across clusters, groups, filaments, nodes, and lower-density regions.

\vspace{-0.2cm}
\subsection{Environmental Characterisation of radio galaxies}
\vspace{-0.2cm}
Environmental characterisation of RGs requires linking radio observables to the physical and large-scale structures in which the sources reside. Host-galaxy properties, group or cluster membership, local galaxy density, distance to filament spines, and location relative to nodes, sheets, and voids all provide complementary measures of environment. Radio properties such as projected size, lobe asymmetry, jet bending, surface brightness, spectral index, and polarisation can then be interpreted in relation to the medium through which the jets and lobes propagate.

The study of radio-galaxy environments has evolved from small targeted samples to large statistical analyses enabled by wide-field radio, optical, and spectroscopic surveys. One of the first quantitative efforts was by \citet{Malarecki2015}, who used spectroscopic data from the 6dF Galaxy Survey and additional redshift compilations to map the three-dimensional galaxy distribution around 19 nearby GRGs (\(z \lesssim 0.2\)). By reconstructing the local density field and identifying filament orientations through positional correlations and inertia-tensor analysis, they showed that these GRGs typically reside in poor groups or field environments. Their radio axes were found to lie preferentially perpendicular to the surrounding filaments, suggesting that jets can expand more efficiently along directions of lower galaxy density. This work provided an early quantitative link between GRG morphology and cosmic-web geometry.

The advent of large-area optical and radio surveys marked a major turning point. Through the SAGAN project, \citet{Dabhade2020a,sagan1} used NVSS and LoTSS data in combination with SDSS and WHL cluster catalogues to show that only \(\sim 15\)--20\% of GRGs reside in clusters, while the majority occupy sparser environments. A follow-up analysis extending to \(z \approx 0.42\) \citep{SAGAN4} further showed that roughly a quarter of GRGs are associated with clusters, with only a small fraction linked to superclusters and most systems lying in lower-density filaments, groups, or field-like environments. More recently, \citet{Mahato2025_SAGANVI} used a filament-centric, fully three-dimensional cosmic-web framework to compare GRGs and SRGs associated with filaments. They found that GRGs are not distinguished from SRGs simply by filament proximity, and that projected size does not show a strong dependence on distance from the filament spine. Instead, FR\,II GRGs show larger jet--filament alignment angles and stronger arm-length asymmetries than FR\,II SRGs, consistent with jet propagation being affected by spine-to-void density gradients and differential ram pressure. These results show that filamentary environments can shape jet propagation and radio morphology.

Wide-field spectroscopic surveys and cosmic-web reconstruction methods now allow RGs to be placed directly within large-scale structure. Using LoTSS DR2 and a T-web reconstruction of the nearby Universe (\(z \lesssim 0.16\)), \citet{Oei2024} carried out a systematic classification of GRGs into filaments, clusters, sheets, and voids. For several thousand RGs, including hundreds of GRGs, they found that \(\sim 60\%\) of GRGs lie in filaments, \(\sim 23\%\) in clusters, \(\sim 17\%\) in sheets, and \(<1\%\) in voids. The strong over-representation of GRGs in filaments and clusters relative to their cosmic volume fractions supports the view that the growth of large radio sources is closely linked to the filamentary structures that channel matter into dense nodes.

These studies show that environmental characterisation is moving from projected associations and small samples towards three-dimensional, cosmic-web-based measurements (see Fig.~\ref{fig:rg_cosmic_web_field} for an illustrative projected view). In the SKA era, such environmental maps will allow radio-galaxy populations to be compared across filaments, groups, clusters, nodes, and lower-density regions, and to test how morphology, projected size, spectral age, and polarisation properties vary with environmental overdensity. Having established how RGs populate the cosmic web, we now turn to how their polarised emission can probe magnetic fields within these large-scale structures.

\vspace{-0.2cm}
\subsection{Polarised Radio Galaxies as Probes of Magnetic Fields in the Cosmic Web}
\vspace{-0.2cm}
As outlined in Sec.~\ref{sec:rgrm}, Faraday rotation offers a direct means of inferring line-of-sight magnetic fields when the polarised emission from a background radio galaxy passes through intervening large-scale structure (see Fig.~D.1 of \citet{SAGAN4}). In practice, however, applying this technique to the cosmic web is non-trivial: it requires accurate three-dimensional localisation of both the polarised source and the foreground structure, careful removal of any local or host-galaxy Faraday contribution, and reliable identification of the relevant filament, group, or supercluster component along the line of sight. Only when this geometry is known with sufficient precision can the measured rotation be converted into a meaningful magnetic-field estimate.

Despite these challenges, recent work has shown the power of this approach. \citet{Sullivan2019} used polarised lobes of background RGs to measure magnetic fields in individual cosmic filaments, while \citet{SAGAN4} and \citet{Pignataro2025} demonstrated that, with well-constrained large-scale structure, magnetic fields of tens of nanogauss can be recovered in the intra-supercluster medium. These results illustrate that polarised RGs can serve as effective beacons for probing magnetisation in distinct parts of the cosmic web, provided their environments and sightlines are robustly characterised.

Looking ahead, wide and dense spectroscopic mapping (e.g., DESI, 4MOST) will supply the necessary three-dimensional structure, while the SKA will provide the high-fidelity RM grids required to extend such measurements systematically. Together, they will enable magnetic-field reconstructions across filaments, groups, and supercluster bridges, opening a pathway to charting the magnetisation of the cosmic web with unprecedented scope and precision.

\vspace{-0.2cm}
\subsection{Radio Galaxies as Probes of Environments}\label{sec:rgprobe}
\vspace{-0.2cm}
RGs provide direct observational constraints on the physical state of the gaseous and magnetised media surrounding AGN. Their large-scale morphologies, spectral properties, and polarisation signatures carry information about gas density, magnetic field strength, and thermal pressure on scales ranging from the circumnuclear ISM to the intracluster medium (ICM).

\subsubsection{Morphology and large-scale environment}
\vspace{-0.2cm}
\begin{itemize}
    \item The shape and extent of radio jets and lobes are sensitive to the external gas distribution. Bent-tail sources such as narrow-angle tail (NAT) and wide-angle tail (WAT) RGs exhibit jet deflections up to $\sim$\,30\deg - 90\deg, produced by ram pressure associated with host-galaxy motion and/or bulk ICM flows in dynamically active group and cluster environments \citep{Begelman1984, OdeaBaumWAT2023}. 
    
    \item These sources trace regions of enhanced ICM density and dynamic cluster environments. Ram pressure models of bent-tail RGs yield typical electron densities of \(n_{\mathrm{e}} \sim 10^{-4}\)–\(10^{-3}\,\mathrm{cm^{-3}}\) in groups and clusters \citep{Freeland2011, Douglass2011, Mao2010}. Observationally, they are reliable signposts of cluster potential wells and merging activity, even at high-$z$ \citep[e.g.,][]{Blanton2003, deVos2021}. 
    
    \item  Interestingly, recent simulations by \citet{Stewart2025} have found that even remnant radio galaxies in dense environments can experience a delay of about 10~Myr in their dynamic transition from active to remnant sources \citep{Walg2014}, after the jets switch-off. Dying, remnant, restarted radio sources have been notoriously difficult to select and characterise, but they are key to quantify the full AGN life cycle, radio-mode AGN feedback at play, and the interaction with the cluster environment \citep[see e.g.,][for a recent review]{Morganti2024}.

    \item FR\,I  RGs are preferentially associated with denser group and cluster environments, whereas FR\,II sources are, on average, more commonly found in lower-density environments, although powerful FR~IIs also occur in clusters and protoclusters \citep{Ineson2013, Ineson2017, Croston2019}.

    Recent LOFAR studies show that the FR\,I/FR\,II transition is not defined by a single luminosity threshold; instead, it spans \(L_{150\,\mathrm{MHz}} \sim 10^{25}\)–\(10^{26}\,\mathrm{W\,Hz^{-1}}\), with morphology governed by host-galaxy mass and environment rather than radio power alone \citep{Mingo2019, Clews2025}.

    \item Giant radio galaxies (GRGs) can serve as in-situ probes of the IGM. Pressure-balance and ram-pressure analyses indicate particle densities of \(n_{\mathrm{IGM}} \simeq (1{-}4)\times10^{-5}\,\mathrm{cm^{-3}}\), at least an order of magnitude lower than those found in rich cluster cores \citep{Mack1998}.

\end{itemize}

A representative example is MSH~05$-$22 (GRG~0503$-$28), a nearby giant radio galaxy (\(z=0.0383\)) hosted by ESO~422$-$G028, with a projected size of \(\sim 1.8\)~Mpc and strong lobe asymmetry, the southern lobe extending \(\sim 1.6\) times farther than the northern one \citep{1986Saripalli}. Early studies showed that the source lies on the edge of a galaxy sheet bordering a void, with its lobes embedded in a warm--hot intergalactic medium (WHIM) whose pressure is comparable to the internal lobe pressure, implying confinement and environmentally driven anisotropic expansion \citep{Subrahmanyan2008}. More recent low-frequency imaging revealed an extended inversion-symmetric X-shaped morphology with a \(\sim 200\)~kpc central radio gap and parallel inner lobe edges over \(\sim 700\)~kpc, interpreted as backflow interaction with the same sheet-/filament-like gaseous structure crossing the source \citep{Dabhade2022msh-xrg}. The correspondence between the inferred WHIM layer and the radio symmetry plane suggests that large-scale gaseous structures can imprint measurable signatures on radio morphology, making MSH~05$-$22 a clear example of how radio continuum mapping can trace the geometry and thermodynamic state of the IGM on megaparsec scales. This example illustrates the kind of asymmetric, megaparsec-scale radio structure for which SKA1-Low, SKA1-Mid, and dense optical/infrared cosmic-web catalogues will make environmental tests routine. By linking jet plasma, diffuse lobes, spectral ageing, and radio morphology to independently mapped filaments, groups, and clusters, SKA-era studies will be able to test both how large-scale structure influences jet propagation and how radio galaxies can act as tracers, and possible agents, of energy and magnetisation within the cosmic web.

\vspace{-0.2cm}
\subsubsection{Polarisation and magneto-ionic environment}
\vspace{-0.2cm}
\begin{itemize}
    \item Linear polarisation and Faraday rotation provide quantitative measures of the magnetised plasma surrounding RGs. Observed RMs range from a few tens to several \(10^{3}\,\mathrm{rad\,m^{-2}}\) in extended lobes, reaching up to \(10^{5}\,\mathrm{rad\,m^{-2}}\) in compact steep-spectrum (CSS) and peaked-spectrum (PS) sources embedded in dense circumnuclear gas \citep{Zavala2004, Saikia2022}.

    \item Asymmetric depolarisation between opposite lobes (the \textit{Laing--Garrington} effect) correlates with jet orientation and foreground plasma density, providing constraints on the distribution and magnetisation of the ICM \citep{Garrington1988, Laing1988}.
\end{itemize}

\vspace{-0.2cm}
\subsubsection{Structural asymmetries and jet--medium interaction}
\vspace{-0.2cm}
Environmental factors such as ICM/IGM pressure, gas motions, local density anisotropy, and magneto-ionic structure can influence jet collimation, bending, confinement, and lobe expansion. Cross-correlated radio and environmental catalogues will allow tests of whether jet brightness, curvature, and lobe-length asymmetries trace local gas-density gradients and anisotropic environments. For example, the BCG-hosted Barbell GRG shows a prominent ($\sim$\,100)-kpc jet kink, possibly linked to plasma instabilities, illustrating why dense cluster environments provide important laboratories for studying complex jet evolution \citep{Dabhade2022Barbell}.

\begin{itemize}
    \item Brightness and length asymmetries between the two lobes can arise from external density and pressure gradients (although orientation and intrinsic jet differences may also contribute). The lobe length ratio in powerful FR\,II sources typically varies by \(10\%\)–\(30\%\) \citep{McCarthy1991, Mullin2008,sagan5}.
    \item Hotspot pressures derived under synchrotron minimum-energy assumptions \citep{Miley1980} are typically of order a few × \(10^{-9}\,\mathrm{dyne\,cm^{-2}}\), reaching up to \(10^{-8}\,\mathrm{dyne\,cm^{-2}}\) in the most powerful FR\,II sources. X-ray analyses of the ambient medium show external pressures that are comparable or slightly lower, indicating approximate pressure balance or mild overpressure at the jet termination regions \citep{Croston2018,Hardcastle2018}.

    \item Equipartition estimates for GRGs yield magnetic field strengths of \(B_{\mathrm{eq}} \sim 0.3{-}1.1~\upmu\mathrm{G}\) and energy densities \(u_{\mathrm{eq}} \sim 10^{-14}\,\mathrm{erg\,cm^{-3}}\). When combined with hotspot cross-sections and lobe-head velocities, these values produce external pressures consistent with the derived IGM densities, confirming that GRGs trace regions of extremely tenuous gas \citep{Mack1998}.

    \item In CSS sources, the detection of strong one-sided depolarisation and high RM values indicates jet propagation through an inhomogeneous, magnetised ISM on sub-kiloparsec scales \citep{OdeaSaikia2021, Saikia2022}.
\end{itemize}

\subsubsection{Feedback and thermodynamic coupling}
\vspace{-0.2cm}
Radio jets and lobes deposit energy, relativistic particles, and magnetic flux into their environments, potentially altering the thermodynamic and magnetised state of the intragroup medium, intracluster medium, and surrounding IGM.
\begin{itemize}
    \item Expanding radio lobes and X-ray cavities displace and heat the surrounding gas, providing a quantitative measure of AGN mechanical feedback and demonstrating that radio jets can regulate the thermal state of the ICM, offset radiative cooling, and influence subsequent star formation.
    \item X-ray observations of clusters such as Perseus, Hydra~A, and MS\,0735.6+7421 show cavities filled with radio plasma carrying energies of \(10^{58}\)–\(10^{61}\,\mathrm{erg}\), sufficient to offset radiative cooling in the ICM \citep{Boehringer1993, Nulsen2005, McNamara2007}.
    \item The measured lobe enthalpy \(4pV\), where \(p\) is the external pressure and \(V\) the cavity volume, combined with cavity ages (\(10^{7}\)–\(10^{8}\,\mathrm{yr}\)), yields typical jet powers of \(10^{43}\)–\(10^{46}\,\mathrm{erg\,s^{-1}}\) \citep{Fabian2012, Croston2011}.

    \item In the FR\,II radio galaxy 3C\,444, \textit{Chandra} imaging reveals a $\sim$200\,kpc shock surrounding the radio structure. The measured temperature jump of a factor of \(\sim1.7\) corresponds to a Mach number \(\mathcal{M} \approx 1.7\), implying an energy deposition of \(E > 8.2\times10^{60}\,\mathrm{erg}\) and jet power \(P_{\mathrm{jet}} > 2.9\times10^{45}\,\mathrm{erg\,s^{-1}}\). Such observations demonstrate that powerful FR\,II jets can drive significant shock heating and mechanical feedback in the ICM \citep{Croston2011}.

    \item Radio-filled X-ray cavities and Faraday-rotation studies indicate that rich-cluster ICM is magnetised to \(B \sim 5{-}40\,\upmu\mathrm{G}\) on coherence scales of \(\sim10{-}50\,\mathrm{kpc}\), with cavity enthalpies \(\gtrsim10^{59}\,\mathrm{erg}\) in systems such as Hydra~A and Perseus. The integrated ICM magnetic energy within \(r \sim 500\,\mathrm{kpc}\) reaches \(\sim1.5\times10^{61}\,\mathrm{erg}\), implying that AGN-inflated, magnetised lobes perform significant \(p\,dV\) work on the hot gas \citep{Taylor1993,Clarke2001,McNamara2000,Fabian2000,Kronberg2001}.

\end{itemize}

\subsubsection{Tracing magnetisation and structure of the intergalactic medium}
\vspace{-0.15cm}
\begin{itemize}
    \item At larger scales, the diffuse lobes of giant RGs and high-redshift FR\,IIs trace the magnetisation of the IGM. Observed inverse-Compton X-ray emission from radio lobes provides estimates of lobe magnetic-field strengths and non-thermal energy densities, enabling comparisons between internal lobe pressure and the surrounding IGM \citep{Croston2005, Ineson2017}.
    
    \item Statistical comparisons of low-frequency radio surveys with cosmological simulations are beginning to connect diffuse synchrotron emission and magnetised plasma with the filamentary cosmic web \citep{Vazza2021,Vernstrom2023}. Combined with cosmic-web classifications of RGs, this provides a route to test whether jets and lobes both trace dense filaments and nodes and contribute to their magnetisation and non-thermal energy budget.

\item Extended and giant RGs store substantial magnetic energy within their radio lobes, with global minimum-energy estimates of \(E_B\sim10^{60}\)–\(10^{61}\) erg in volumes approaching \(10^{73}\,\mathrm{cm^3}\) \citep{Kronberg2001}. For the low-frequency polarised GRGs detected by LOFAR, depolarisation modelling indicates rarefied local environments with \(n_{\rm e}<10^{-5}\,\mathrm{cm^{-3}}\), magnetic fields \(B<0.1\,\mu\mathrm{G}\), and field fluctuations on scales of approximately \(3\)–\(25\) kpc \citep{Stuardi2020}. The subsequent expansion and mixing of their lobes may distribute magnetised plasma over megaparsec scales, although the efficiency and volume filling factor of this process remain uncertain.

\end{itemize}
These examples show that radio lobes are both probes and agents of environmental evolution, tracing gas pressure, density structure, and magneto-ionic conditions while also depositing energy, relativistic particles, and magnetic flux into the ICM and IGM.

\vspace{-0.2cm}
\subsubsection{Jet energetics and environmental diagnostics}
Powerful FR\,II RGs serve as quantitative diagnostics of their gaseous environments. Using a self-similar lobe expansion model constrained by ram-pressure balance at the jet head, \citet{Daly1995} derived expressions linking observable radio parameters to intrinsic physical quantities such as the ambient gas density, jet kinetic power, and source age. Application of this framework yielded typical external densities of \(n_{\mathrm{ext}} \sim 10^{-4}\)–\(10^{-2}\,\mathrm{cm^{-3}}\) at radii of 50-100\,kpc, jet powers of \(L_{\mathrm{j}} \sim 10^{44}\)–\(10^{46}\,\mathrm{erg\,s^{-1}}\), and dynamical ages of \(t \sim 10^{7}\)–\(10^{8}\,\mathrm{yr}\). The resulting lobe pressures, \(p_{\mathrm{lobe}} \sim \text{few} \times 10^{-10}\)–\(10^{-9}\,\mathrm{dyne\,cm^{-2}}\), exceeded the surrounding thermal pressure by factors of a few, implying mildly supersonic expansion and ongoing work done on the external medium. 

Subsequent radio–X-ray studies with \textit{Chandra} and \textit{XMM–Newton} \citep[e.g.][]{Croston2005,Croston2011,Croston2018} directly measured both the lobe and external gas pressures, empirically validating and refining these estimates. They found that FR\,II lobes are typically close to pressure balance or modestly overpressured, whereas FR\,I lobes tend to be underpressured and particle-dominated. Inverse-Compton detections of extended lobes further indicated magnetic field strengths of \(B \approx 0.3\)–\(1\,B_{\mathrm{eq}}\), consistent with near-equipartition conditions. Together, these results show that extended RGs trace the thermodynamic state of their environments and can provide model-dependent constraints on ambient density, pressure, and jet energy injection, including where direct X-ray measurements are weak or unavailable. Such diagnostics connect the microphysics of relativistic jets to the macroscopic thermodynamics of the IGM and ICM, although the inferred quantities depend on assumptions about source geometry, particle content, magnetic field strength, filling factor, and dynamical state.

The broad frequency leverage of SKA1-Low and SKA1-Mid will enable spectral-index and spectral-curvature measurements for large RG samples, helping to distinguish aged, remnant, or restarted plasma from young and actively fuelled jets. Such measurements will test whether spectral age, remnant fraction, restarted activity, and lobe expansion depend on environmental confinement, density, and pressure.

\vspace{-0.3cm}
\section{Studying the High-Redshift Cosmic Web with Protoclusters and Radio Galaxies} 
\vspace{-0.2cm}
\subsection{Galaxy Protoclusters} \label{sec:pc}
\vspace{-0.2cm}
Galaxy protoclusters represent the largest non-virialised overdensities in the early Universe, tracing the high peaks of the primordial density field that will collapse into present-day clusters with masses of \(M_{z=0} \gtrsim 10^{14}\,M_\odot\) \citep[see e.g.][for some reviews]{Overzier2016,Alberts2022}. Rather than being uniformly filamentary objects, protoclusters are extended, multi-halo overdensities embedded within the cosmic web. They are commonly identified at $z\gtrsim2$, although the protocluster phase is not defined by a strict redshift threshold. Indeed, in the hierarchical structure-formation framework, protoclusters are spatially extended overdense regions spanning tens of comoving Mpc at high redshift and evolving into gravitationally bound clusters by $z=0$ \citep{Chiang2013,Muldrew2015}.

Protoclusters serve as laboratories for studying the emergence of environmental effects and the growth of massive structures. They host intense star formation and AGN activity, accounting for a disproportionately high fraction of the cosmic star-formation-rate density during the first two billion years \citep{Chiang2017}. Recent spectroscopic surveys have revealed exceptional examples, such as the Taralay protocluster at \(z\simeq4.57\) (\(M_{z=0}\simeq1.7\times10^{15}\,M_\odot\)) and SXDS\_gPC at \(z\simeq5.7\), which exhibits a Ly\(\alpha\)-emitter overdensity of \(\delta_g\simeq5.6\pm1.2\) \citep{Jiang2018,Staab2024}. The latter is projected to evolve into a cluster of \(M_{z=0}\!\approx\!(3.6\pm0.9)\times10^{15}\,M_\odot\), comparable to the most massive known clusters. Such giant protoclusters are extremely rare and cosmological simulations show that the likelihood of finding one within a few square degrees at \(z\!>\!5\) is only a few per cent \citep{Chiang2013, Overzier2016}. Their discovery thus provides stringent constraints on models of structure growth and primordial non-Gaussianity.

\subsection{High-redshift radio sources as tracers of protoclusters}\label{sec:highzrg}
\vspace{-0.2cm}
High-redshift RGs (HzRGs) and radio-loud quasars (HzRQs) are among the most powerful beacons of protocluster environments \citep[e.g.,][]{Miley2008,Hatch2014}. Notable examples include MRC~1138-262, also known as the Spiderweb Galaxy, a powerful RG at $z=2.16$ hosted by a rich protocluster showing intense star formation and merging activity \citep{Dannerbauer2017,Jin2021,Shimakawa2025}. Other HzRGs associated with protoclusters include TN~J1338--1942 ($z=4.1$), embedded in an overdensity containing numerous Ly$\alpha$-emitting galaxies \citep{Intema2006,Duncan2023}; 4C~41.17 and 4C~60.07, both at $z=3.8$, associated with a protocluster rich in submillimetre galaxies \citep{Ivison2000,DeBreuck2004}; USS~1558-003 ($z=2.53$), surrounded by H$\alpha$ emitters and submillimetre sources \citep{PerezMartinez2024,Tadaki2014}; TGSS~J1530+1049 at $z=4.0$, located in an H$\alpha$-emitter overdensity \citep{Saxena2026}; and 7C~1756+6520 ($z\sim1.4$), which hosts a protocluster with a high AGN fraction ($\sim23\%$) and a molecular-gas-rich AGN member \citep{Casasola2013,Casasola2018}.

HzRGs are typically hosted by the most massive galaxies (\(M_{\ast}\!\sim\!10^{11}\!-\!10^{12}\,M_\odot\)) in the early Universe and exhibit large-scale overdensities of star-forming companions. Systematic studies show that \(>70\%\) of RGs at \(2\!\lesssim\!z\!\lesssim\!5\) lie within LAE or H$\alpha$-emitter overdensities consistent with protoclusters \citep{Venemans2007, Hatch2011a, Cooke2014}. \citet{Castignani2014} studied a sample of low-luminosity $z\sim1-2$ RGs, drawn from the COSMOS survey.
They found that $\sim70\%$ of them are in dense Mpc-scale environments, similarly to previous lower-redshift studies \citep{Zirbel1997}. As low-luminosity RGs represent the bulk of the RG population, these results have implications for the search for distant clusters and protoclusters in ongoing and forthcoming wide-field surveys such as {\it Euclid} and LSST, using RGs as tracers. The CARLA survey found comparable, though somewhat lower, overdensity fractions for powerful radio-loud AGN: 55\% of 387 fields at \(1.3<z<3.2\) are overdense at the \(2\sigma\) level, with similar environments for RGs and RQs \citep{Wylezalek2013,Hatch2014}. Such dense environments appear to favour jet formation and sustained AGN feedback \citep{Hatch2014}. In turn, powerful RGs can trigger or quench star formation through jet-induced shocks and mechanical feedback, as observed in systems such as the Spiderweb Galaxy at \(z=2.2\) \citep{Nesvadba2006,Miley2006}. These sources therefore mark the nodes of the forming cosmic web, illuminating gas-rich filaments and pre-virialised haloes through radio, Ly\(\alpha\), and X-ray emission.

The interaction between RG jets and dense environments can strongly influence radio morphology. At low and intermediate redshifts, tailed RGs are commonly found in clusters, often in association with the brightest cluster galaxies \citep{Owen_Rudnick1976,deVos2021}; LOFAR and other SKA precursors have been pivotal in characterising such sources and quantifying environmental effects on size, lobe symmetry, and radio luminosity \citep{GoldenMarx2023}. Young and compact RGs, including compact steep-spectrum and GHz-peaked-spectrum sources, have also been reported in cluster environments \citep{deVries2000,Drake2004,OdeaSaikia2021}. With its sensitivity, angular resolution, and wide frequency coverage, the SKA will extend these studies to large samples of RGs in protoclusters, spanning a broad range of morphologies. It will also enable spectral-ageing analyses of the cores, jets, and lobes of distant RGs, providing deeper insight into the morphology, evolution, and particle-acceleration properties of HzRGs \citep[see also ][]{Afonso01.2026.SKA}.

Furthermore, in the \textit{SKA era}, deep continuum and polarisation mapping combined with spectroscopic redshifts will enable a large, statistically well-defined census of RGs and RQs in protoclusters. Their polarised synchrotron emission will trace magnetised intergalactic plasma and AGN-driven outflows, allowing direct tests of feedback, magnetisation, and gas accretion within assembling cluster environments across cosmic time.

\begin{figure}
    \centering
	\includegraphics[width=\columnwidth]{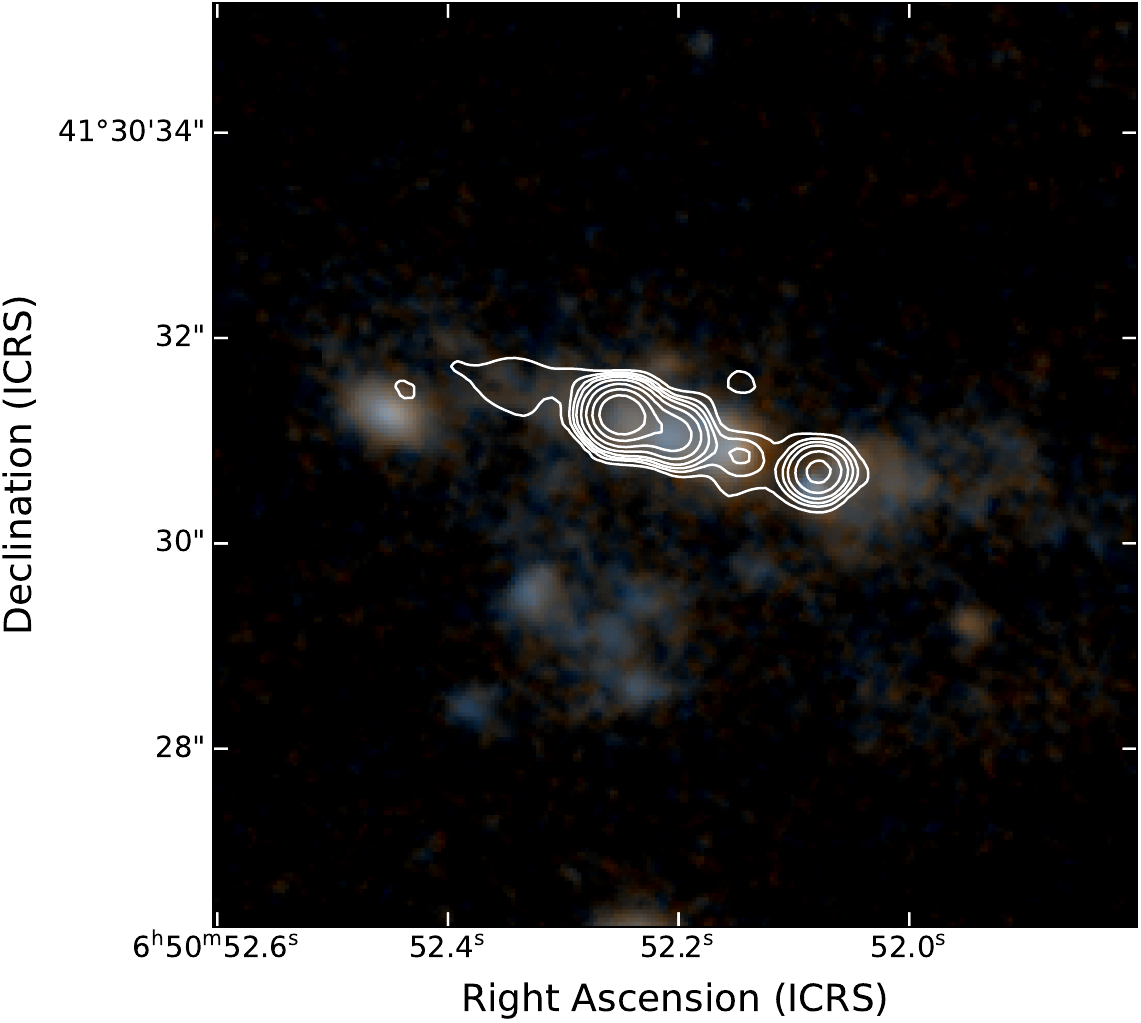}
    \caption{4C~41.17 ($z\sim3.8$): Hubble Space Telescope WFC3/IR colour composite image of the field centred at RA = 102.721 and Dec = +41.510 (ICRS), constructed from F105W and F160W filter data. The image combines 16 exposures (total integration time $\approx$ 653~s) with a pixel scale of 0.04 arcsec/pixel. North is up, and east is to the left.  White contours represent \textit{VLA} 4.71~GHz emission from the NRAO VLA Archive Survey (NVAS), 
    plotted at $3\sigma \times [1,\,2.3,\,3.5,\,5.2,\,8.5,\,14.5,\,25,\,40,\,76.5]$, where 
    $\sigma = 0.0002$~Jy~beam$^{-1}$. The synthesised beam is 0.41\arcsec $\times$ 0.37\arcsec, -84.9\deg. For detailed discussion, see Sec.~\ref{sec:highzrg} \& Sec.~\ref{sec:highzmag}.}
    \label{fig:4C41.17}
\end{figure}

\textit{4C~41.17: A Radio Galaxy in a Forming Protocluster}:\\
One of the most compelling examples of an RG associated with a forming protocluster is 4C~41.17 (see Fig.~\ref{fig:4C41.17}), at $z = 3.792$--3.796 \citep{DeBreuck2001}. It was among the first HzRGs identified with an extended Ly$\alpha$ halo of nearly 200~kpc and a strong radio–optical alignment over $\sim70$~kpc. High-resolution \textit{Hubble Space Telescope} (HST) and Keck imaging revealed multiple UV and optical knots aligned with the radio axis, consistent with jet-induced star formation, while the UV continuum shows 10–20\% polarisation, indicating a combination of scattered AGN light and young stellar emission \citep{Dey1997}. The Ly$\alpha$ nebula, with $L_{\mathrm{Ly}\alpha} \sim 10^{44}$~erg\,s$^{-1}$ and a velocity gradient of $\sim600$~km\,s$^{-1}$, traces large-scale gas motions within the forming environment \citep{Reuland2003}.

The host harbours a powerful AGN embedded in an ultraluminous infrared galaxy with $L_{\mathrm{FIR}} \simeq (1.5 \pm 0.3)\times10^{13}\,L_\odot$ and $T_{\mathrm{dust}} \approx 54 \pm 10$~K, consistent with an intense starburst \citep{Ivison2012}. The system contains $M_{\mathrm{dust}} \sim 1.5\times10^8\,M_\odot$ and a stellar mass $M_\ast \sim 10^{11}\,M_\odot$ \citep{DeBreuck2005}. CO(4--3) observations reveal two gas components separated by $1.8''$ ($\approx13$~kpc) and $\sim400$~km\,s$^{-1}$, each with $M_{\mathrm{H_2}} \approx 3\times10^{10}\,M_\odot$, implying a dynamical mass $\gtrsim6\times10^{10}\,M_\odot$ and a systemic redshift of $z_{\mathrm{CO}} = 3.7958 \pm 0.0008$ \citep{DeBreuck2005}.

Deep submillimetre mapping uncovered a surface density approximately ten times the field value of bright SMGs within $\sim2.5'$ ($\sim1$~Mpc), with a probability of $P \approx 1.7\times10^{-3}$ that this is a chance alignment \citep{Ivison2000}. Many of these SMGs have $S_{850} \lesssim 16$~mJy and very red colours ($V-K \gtrsim 7$), characteristic of massive, dusty starbursts. Subsequent \textit{Spitzer} and \textit{Herschel} observations confirmed significant mid-infrared overdensities \citep{Wylezalek2013} and identified at least one bright far-IR source within $\sim25''$ sharing the radio galaxy’s redshift \citep{Greve2007}. 

X-ray observations with \textit{Chandra} revealed both a luminous nuclear component ($L_{2-10\,\mathrm{keV}} \sim 10^{45}$~erg\,s$^{-1}$) and extended soft X-ray emission aligned with the radio axis over $\sim100$~kpc \citep{Scharf2003, Overzier2005}. The diffuse emission is attributed to inverse-Compton scattering of cosmic microwave background photons by the radio lobes, possibly augmented by shocked or heated gas from early intracluster medium formation. 

Together, the extended ionised halo, powerful AGN activity, diffuse X-ray plasma, large molecular gas reservoirs, and surrounding dusty starbursts within $\sim1$~Mpc establish 4C~41.17 as a dense node of galaxy formation and a well-studied example of a massive protocluster core in rapid assembly at $z \approx 3.8$. It exemplifies how high-redshift RGs can serve as signposts of forming cluster environments, where black hole growth, star formation, and hot gas buildup are simultaneously shaping the early stages of massive structure formation.

\subsection{Multi-wavelength Synergies for Studying Protoclusters in the High-Redshift Universe}
\vspace{-0.2cm}
Building on previous searches for distant clusters associated with low-power RGs \citep[e.g.,][]{Castignani2014}, which constitute the majority of the RG population, and taking into account the steepness of the cluster mass function \citep{Bode2001}, we expect the discovery of $\sim16,000$ clusters around RGs at $z \sim 1$--2 (typically coinciding with the brightest cluster galaxies), with signal-to-noise ratios exceeding $S/N > 3$ and down to cluster masses of $\sim2\times10^{14}\,M_\odot$, by the time both the \textit{Euclid} and SKA observatories are operational. This forecast corresponds to a completeness of $\sim90\%$ \citep{Sartoris2016,Adam2019}, comparable to that of the best current samples of distant clusters. 
Radio-loud BCGs and their associated distant (proto)clusters, to be jointly identified by the SKA and \textit{Euclid} surveys \citep[e.g.][]{Prandoni2015,Bohringer2025}, will vastly outnumber all existing samples, offering an unprecedented window on the formation and evolution of large-scale structures and radio galaxies, and their interplay through radio-mode AGN feedback. The SKA will not only detect radio galaxies at the centres of distant (proto)clusters but will also provide high-sensitivity maps of the surrounding environments. With its wide frequency coverage (50\,MHz–15\,GHz) and unprecedented continuum sensitivity of $\lesssim1\,\mu$Jy, the SKA will enable the detection of star-forming galaxies both in protocluster cores \citep{Daddi2017} and along their extended filamentary structures \citep{Jin2021}, in continuum emission up to $z \sim 2$ and beyond, and in H\,\textsc{i} line emission up to $z \sim 1$.

\textit{JWST} is rapidly transforming the identification and physical characterisation of protoclusters and overdense galaxy environments across the early Universe. Its NIRCam and NIRSpec capabilities have enabled spectroscopic confirmation or identification of structures at \(z>7\), including A2744-z7p9OD at \(z=7.88\) \citep{Morishita2023}, a lensed protocluster candidate behind SMACS0723 at \(z=7.66\) \citep{Laporte2022}, and overdense environments in the JADES fields \citep{Li2026}. These very high-redshift systems demonstrate JWST's ability to map structure formation close to the epoch of reionisation, while ongoing and forthcoming JWST programmes at \(z\sim2\)--5 will provide much more detailed views of the dense environments in which massive galaxies, AGN, and intracluster structures are rapidly assembling. This makes the SKA particularly timely: JWST can define the stellar, nebular, and environmental properties of forming structures, while SKA observations can search for radio-loud AGN, compact jets, diffuse synchrotron emission, and early radio-mode feedback within the same environments \citep[see also ][]{Mazzolari01.2026.SKA}. Existing JWST studies of HzRG protoclusters, including TN~J1338-1942 at \(z\simeq4.1\) and the Spiderweb protocluster at \(z=2.16\), already illustrate the power of combining rest-frame optical/near-infrared diagnostics with radio-selected tracers of dense forming structures \citep{Duncan2023,Shimakawa2025}.

\subsection{Probing Cosmic Magnetism in the High-Redshift Universe}\label{sec:highzmag}
\vspace{-0.2cm}
The SKA, with its unprecedented sensitivity and resolution, will uncover and resolve a large number of HzRGs, many embedded within forming protoclusters. The simulations of \citet{Mazzolari2024} forecast that SKA radio surveys will provide a dust-unbiased route to early AGN populations, with \(>2000\) AGN predicted at \(z>6\) and several tens at \(z>10\), including a substantial Compton-thick fraction. These forecasts should be treated as predictions for high-redshift radio-selected AGN; confirming them as classical HzRGs or protocluster members will require host identifications, redshifts, and environmental mapping, especially as extended lobes may be CMB-suppressed at the highest redshifts \citep[e.g.,][]{Ghisellini2015}.

The study of HzRG--protocluster systems is valuable because joint radio, X-ray, submillimetre, and SZ observations can constrain the magnetised plasma within radio lobes and its coupling to the surrounding hot gas. One well-studied case is 4C~41.17. The diffuse X-ray emission around 4C~41.17 broadly follows the radio morphology and has a non-thermal spectrum consistent with inverse-Compton scattering by relativistic electrons associated with the radio source \citep{Scharf2003}. In general, comparison of the radio synchrotron and inverse-Compton X-ray emission provides an independent constraint on the lobe magnetic field when the energy density of the seed-photon field is known. For samples of $z\sim2$ radio galaxies in which IC/CMB emission is expected to dominate, this method yields magnetic-field strengths of $B_{\rm IC}\sim30$--$180\,\mu\mathrm{G}$, broadly consistent with equipartition estimates of $B_{\rm eq}\sim100$--$200\,\mu\mathrm{G}$ \citep{Overzier2005}. For 4C~41.17, the contribution of a luminous, spatially extended far-infrared photon field associated with the source must also be considered. At $z=3.8$, the CMB energy density is $U_{\rm CMB}\simeq2.2\times10^{-10}\,\mathrm{erg\,cm^{-3}}$. Adopting the equipartition field of $B_{\rm eq}\simeq50$--$80\,\mu\mathrm{G}$ inferred for the low-surface-brightness radio regions, \citet{Scharf2003} argued that the CMB alone could account for at most $\sim40\%$ of the observed inverse-Compton emission, with the local far-infrared radiation field providing additional seed photons. The relative contributions of the CMB and local far-infrared emission nevertheless remain model-dependent, owing to uncertainties in the magnetic field, electron-energy distribution, and photon-field geometry.
Consequently, 4C~41.17 illustrates how combined radio synchrotron and IC X-ray measurements can constrain the magnetic field and relativistic-particle content of high-redshift radio lobes. If the lobes are approximately in pressure balance with their surroundings, these measurements, together with dynamical modelling and X-ray or SZ constraints, can also place indirect limits on the density and total pressure of the ambient protocluster medium. Separating the external magnetic-field contribution, however, requires additional Faraday-rotation or depolarisation measurements and an independent constraint on the electron density.

Recent multi-wavelength observations of the Spiderweb protocluster (PKS 1138-262 at $z = 2.16$) demonstrate the power of combining radio, X-ray, and millimetre data to probe the co-evolution of magnetic and thermal energy in forming clusters. A 6$\sigma$ detection of the thermal SZ effect with ALMA and ACA revealed a nascent intracluster medium with $M_{500} \simeq (3$--$8)\times10^{13}\,M_\odot$ extending over $r_{500} \sim 250$~kpc \citep{DiMascolo2023}. The system shows spatial offsets between the SZ, X-ray, and radio emission, consistent with an unrelaxed, multi-phase environment shaped by powerful AGN feedback. The non-thermal energy of the radio jet ($\sim6\times10^{60}$~erg) is comparable to the thermal energy of the surrounding gas, implying that magnetic fields and relativistic plasma already play a significant role in structuring the early intracluster medium. When combined with radio and X-ray data, SZ measurements add the crucial thermal perspective by quantifying the pressure and energy content of the hot gas. This enables a direct comparison between the non-thermal (radio and IC X-ray) and thermal (SZ) energy reservoirs, revealing how efficiently AGN jets heat and magnetise their environments. Therefore, it provides a benchmark for future SKA-era studies of the growth of magnetised structure across cosmic time.

Together with sources such as 4C~41.17, the Spiderweb system highlights how RGs trace the emergence of magnetised baryonic structures in the high-redshift Universe. In the \textit{SKA era}, deep radio polarimetric and continuum surveys combined with X-ray, submillimetre, and SZ data will enable magnetic-field strengths and non-thermal energetics to be constrained in selected protocluster environments across a broad redshift range, allowing the evolution of cosmic magnetism to be traced region by region and epoch by epoch.

In protocluster environments at \(z>3\), structures such as SXDS\_gPC \citep{Jiang2018}, with galaxy overdensities of \(\delta_g \sim 5\) and predicted present-day masses exceeding \(3\times10^{15}\,M_\odot\), illustrate the kinds of forming dense environments in which SKA observations can search for early radio-mode feedback, pre-heating, and magnetic enrichment of the proto-ICM.

\subsection{Tracing Star Formation in Protoclusters Through Radio Emission}
\vspace{-0.2cm}
A complete view of radio-galaxy protocluster environments requires tracing not only the central AGN and magnetised plasma, but also the build-up of stellar mass in the surrounding galaxy population. This is a key component of the baryon cycle in forming clusters, where gas accretion, star formation, AGN activity, and feedback together regulate how baryons are converted into stars, heated, expelled, or retained during cluster assembly. In radio-loud AGN, the radio continuum is dominated by jet-, core-, and lobe-related synchrotron emission, and therefore cannot generally be used as a direct star-formation-rate tracer without careful AGN subtraction. For other protocluster members whose radio emission is not AGN-dominated, however, radio continuum emission provides a dust-independent measure of star formation, tracing both synchrotron radiation from supernova-accelerated cosmic rays and thermal free-free emission from H\,\textsc{ii} regions. Through the far-infrared--radio correlation \citep{Condon1992,Yun2001,Bell2003}, radio luminosity can then be used to map obscured star formation across protocluster cores, outskirts, and filamentary structures.

\subsubsection{Current state of the art: constraints from deep LOFAR surveys}
\vspace{-0.2cm}
Deep LOFAR surveys now provide the empirical basis for radio--SFR work at frequencies relevant to SKA1-Low. The LoTSS Deep Fields (ELAIS-N1, Boötes, and Lockman Hole) cover \(\sim26\,\mathrm{deg^2}\), with \(>100\)~h integrations per field, \(\sim20\,\upmu\mathrm{Jy\,beam^{-1}}\) rms sensitivity, and \(6\arcsec\) resolution. Their extensive UV--to--FIR coverage enables robust AGN/star-forming galaxy separation and 150\,MHz SFR calibration: \citet{Best2023} derived \(\log L_{150}=22.24+1.08\,\log(\mathrm{SFR})\), consistent within \(\sim0.1\) dex with \citet{Smith2021}. The intrinsic scatter is \(\sim0.3\) dex and, if uncorrected, biases the inferred SFR density by \(\sim4\%\), as quantified by \citet{Cochrane2023}. These results show that low-frequency radio surveys can recover the cosmic star-formation history to \(z\sim4\) with precision comparable to far-infrared studies, while avoiding dust attenuation and providing a direct empirical framework for SKA continuum SFR studies. The interpretation nevertheless remains sensitive to AGN contamination, intrinsic scatter and possible evolution in the radio--SFR relation, spectral-index and \textit{K}-correction assumptions, and inverse-Compton losses at high redshift.

\subsubsection{Probing star formation in protoclusters with the SKA}
Protoclusters are dense, rapidly evolving environments where galaxies assemble, and star formation is often heavily obscured. Optical and ultraviolet tracers can miss a substantial fraction of this activity, while infrared measurements may be limited by confusion and source blending in crowded fields. For protocluster members whose radio emission is not AGN-dominated, SKA radio continuum observations will provide a dust-insensitive route to mapping star formation across protocluster cores, outskirts, and filaments.

At high rest-frame frequencies, radio emission becomes increasingly dominated by thermal free--free radiation, which provides a more direct tracer of the ionising photon rate. This is particularly valuable at high redshift, where synchrotron emission is increasingly suppressed by inverse-Compton losses against the CMB, whose energy density scales as \((1+z)^4\). Observations near 10 GHz with SKA1-Mid Band~5 will probe rest-frame frequencies of approximately \(30\)–\(60\) GHz over \(z\sim2\)–5. Under the ultra-deep, \(\sim1000\)-h reference-survey assumptions considered by \citet{Murphy2015}, galaxies with SFRs of order \(\sim100\,M_\odot\,\mathrm{yr^{-1}}\) are expected to be detectable across a broad redshift range. This represents an ultra-deep survey benchmark rather than the sensitivity of a typical targeted pointing.

A tiered SKA strategy will therefore be well matched to this science: wide surveys can identify rare massive structures, deep observations of known protoclusters can reach the active star-forming population at \(z\sim3\)--5, and ultra-deep fields can probe the earliest systems at \(z>4\). Combining low- and high-frequency SKA data will separate synchrotron and thermal components, measure spectral indices, and test whether star formation in protoclusters is governed mainly by local galaxy properties or by the surrounding large-scale environment \citep[see also ][]{Algera01.2026.SKA,FangxiaAn01.2026.SKA}.

\vspace{-0.35cm}
\section{Observational Strategy and Quantitative Expectations}
\vspace{-0.3cm}
The advent of the SKA marks a decisive transition in our ability to study RGs within the evolving large-scale structure of the Universe. The SKA capabilities introduced in Section~\ref{sec:ska_framework} motivate a tiered observational strategy for using RGs as probes of the cosmic web. With representative continuum sensitivities of \(\sim1\)--\(2\,\upmu\mathrm{Jy\,beam^{-1}}\) at 1.4\,GHz and tens of \(\upmu\mathrm{Jy\,beam^{-1}}\) at SKA1-Low frequencies, SKA1 surveys will sample radio-source populations over a wide range of luminosities, from low-power FR\,I systems in groups to powerful FR\,II sources and GRGs extending beyond 1\,Mpc. Wide-area surveys will provide the statistical samples needed to compare radio-source populations across clusters, groups, filaments, sheets, voids, and protocluster environments, while deeper fields will be required to detect faint, low-surface-brightness, remnant, restarted, and high-redshift systems.

\subsection{Science Drivers and Physical Probes for Cosmic-Web Studies with the SKA}
\vspace{-0.2cm}
The framework proposed here is to use SKA observations to turn radio galaxies from isolated radio detections into quantitative probes of cosmic-web environments. This requires an observing strategy capable of recovering the relevant radio plasma, and an interpretation strategy that links those measurements to independently mapped large-scale structure. The key question is not simply whether a radio galaxy is detected, but whether its physical structure and plasma evolution can be measured: compact cores and inner jets require high angular resolution, extended lobes require adequate surface-brightness sensitivity, and remnant or restarted plasma requires low-frequency coverage. The angular-resolution component of this framework is illustrated in Fig.~\ref{fig:ska_morph_resolvability}. No single SKA observing mode will meet all these requirements. SKA1-Mid will resolve compact structures and tens-of-kpc radio galaxies across much of cosmic time, whereas SKA1-Low will provide the low-frequency sensitivity and survey capability required to recover aged, diffuse, steep-spectrum plasma on scales of hundreds of kpc to several Mpc. Wide surveys are needed to build statistical samples across clusters, groups, filaments, sheets, voids, and protoclusters, while deeper fields are required to detect faint, diffuse, remnant, restarted, and high-redshift populations. These radio measurements can then be combined with host identifications, redshifts, and optical, infrared, X-ray, and SZ cosmic-web catalogues to test how environment regulates radio-galaxy evolution and how jets and lobes deposit energy and magnetic flux into the surrounding medium.

The quantitative scale of this programme is set by both SKA forecasts and SKA pathfinder results. Forecasts for an SKA1-Mid all-sky continuum survey such as SASS1 predict \(\sim5\times10^{8}\) radio sources at \(\sim2^{\prime\prime}\) resolution and \(2~\upmu\mathrm{Jy~beam^{-1}}\) rms \citep{sass12015Norris}. Recent EMU results already demonstrate that automated radio-galaxy cataloguing can operate at the required scale. In the \(270~\mathrm{deg^2}\) EMU pilot survey, the RG-CAT pipeline produced \(211{,}625\) radio sources, including \(10{,}414\) extended radio sources \citep{Gupta2024RGCAT}. The same work achieved \(\mathrm{IoU}>0.5\) for 99\% of central radio-galaxy bounding boxes and placed 98\% of predicted host positions within \(3^{\prime\prime}\) of the ground-truth CatWISE host. More recently, EMUSE applied morphology-based retrieval to first-year EMU main-survey data, covering \(\sim4{,}500~\mathrm{deg^2}\), with \(\sim3\times10^{6}\) detected radio sources and \(\sim1.7\times10^{5}\) extended radio sources suitable for similarity search \citep{Gupta2025EMUSE}. These precursor results show that a conservative SKA-era expectation of \(>10^{5}\) morphologically measurable radio galaxies is well grounded; first-year EMU-scale data already reach this order of magnitude for extended-source samples, and SKA depth, resolution, and sky coverage should expand the accessible population substantially. Citizen-science programmes could provide a complementary route for inspecting complex, rare, or ambiguous radio morphologies, while also generating validated training and test samples for increasingly automated machine-learning classification \citep{Hota01.2026.SKA}.

Within this framework, the key SKA observables can be linked directly to physical diagnostics:
\vspace{-0.2cm}
\begin{itemize}
    \item Morphology and size: Angular size, projected linear size, arm-length ratio, lobe axial ratio, core dominance, and misalignment angle will trace jet propagation, environmental asymmetry, confinement, and the growth of radio plasma. Fig.~\ref{fig:ska_morph_resolvability} quantifies the practical resolution thresholds: under the adopted five-beam criterion, SKA1-Mid at 1.4\,GHz can resolve radio structures larger than \(\sim10\,\mathrm{kpc}\) across much of cosmic time, while SKA1-Mid at 6.7\,GHz can resolve structures of a few kiloparsecs. SKA1-Low at 158\,MHz is instead most effective for structures of several hundred kiloparsecs, making it especially suited to diffuse lobes, remnant plasma, and giant radio galaxies.

    \item Spectral state and ageing: Multi-frequency SKA observations will measure spectral index, curvature, and spectral ageing, separating young, active, remnant, restarted, and diffuse plasma. These diagnostics will test whether dense environments prolong radio-source visibility, suppress lobe expansion, or favour recurrent jet activity.
    
    \item Polarisation and Faraday rotation: Current ASKAP pathfinder surveys already achieve median RM uncertainties of order \(2\,\mathrm{rad\,m^{-2}}\) over most of the sky, with SPICE-RACS DR2 providing \(\sim2.5\times10^{5}\) reliable RMs at an areal density of \(\sim6.7\,\mathrm{deg^{-2}}\) \citep{Thomson2026SPICE-RACS}. The SKA will greatly increase the source density and angular sampling required to constrain magnetic-field strength, coherence, and topology across radio galaxies, clusters, filaments, and other cosmic-web environments.

    \item Environment and Feedback: Cross-matching with DESI, 4MOST, \textit{Euclid}, LSST, X-ray, and SZ surveys will provide host identifications, redshifts, overdensities, web-type classifications, and ambient thermal pressures. These data will allow radio luminosities, lobe pressures, asymmetries, and magnetic signatures to be tested against the environment, enabling quantitative constraints on energy injection and magnetic enrichment of the ICM and IGM.

\end{itemize}

The central goal is therefore not simply to detect more radio galaxies, but to build statistically controlled samples in which radio morphology, spectra, polarisation, host properties, and cosmic-web location are measured together. This will allow SKA surveys to test how large-scale environment regulates radio-AGN life cycles, how jets and lobes deposit energy and magnetic flux into their surroundings, and how radio galaxies trace the baryonic and magnetised structure of the Universe across cosmic time.

\vspace{-0.35cm}
\section{Summary}
\vspace{-0.2cm}
In this chapter, we have reviewed how radio galaxies trace, respond to, and influence the large-scale structure of the Universe, and outlined how the SKA, together with complementary multi-wavelength facilities, can turn these connections into quantitative and testable probes of cosmic-web evolution.

Over the past decades, radio surveys, spectroscopic mapping, and cosmological simulations have revealed that radio galaxies carry far richer environmental information than once appreciated. Their morphology, spectral state, and magnetised plasma respond to density gradients, gas flows, and magnetic-field structure across the cosmic web. Present constraints, however, remain limited by sample size, polarimetric depth, selection effects, and the difficulty of locating each source precisely within its three-dimensional environment.

The sensitivity, resolution, survey speed, and polarisation capabilities of the SKA will substantially reduce these limitations and enable statistically controlled studies of radio galaxies across the full range of cosmic-web environments. By combining radio morphology, spectral and Faraday information, environmental context, and redshift measurements, SKA surveys will test how feedback influences filament and cluster thermodynamics, when intergalactic magnetic fields emerge, and whether galaxy and black-hole growth depends on cosmic-web location. The SKA will therefore transform radio galaxies from largely qualitative signposts into quantitative probes of structure formation.

In the \textit{SKA era}, radio galaxies will no longer simply inhabit the cosmic web; they will decode it. And in doing so, they will allow us to read, for the first time, the magnetic, dynamical, and evolutionary script of the Universe itself.

\vspace{-0.35cm}
\section{Acknowledgements}
SS acknowledges the support from the Estonian Research Council grant PSG1045. We acknowledge that this work has made use of  \textsc{astropy} \citep{astropy} and \textsc{aplpy} \citep{apl}.

\bibliographystyle{abbrvnat-maxbibnames4}

\bibliography{chapter} % if your bibtex file is called example.bib

\end{document}